\documentclass[%
 preprint,
 nofootinbib,
 amsmath,amssymb,
 aps,
tightenlines]{revtex4-1}

\usepackage{graphicx}
\usepackage{dcolumn}
\usepackage{bm}

\usepackage{mathrsfs}

\begin{document}

\def\a{\alpha}
\def\b{\beta}
\def\c{\varepsilon}
\def\d{\delta}
\def\e{\epsilon}
\def\f{\phi}
\def\g{\gamma}
\def\h{\theta}
\def\k{\kappa}
\def\l{\lambda}
\def\m{\mu}
\def\n{\nu}
\def\p{\psi}
\def\q{\partial}
\def\r{\rho}
\def\s{\sigma}
\def\t{\tau}
\def\u{\upsilon}
\def\v{\varphi}
\def\w{\omega}
\def\x{\xi}
\def\y{\eta}
\def\z{\zeta}
\def\D{{\mit \Delta}}
\def\G{\Gamma}
\def\H{\Theta}
\def\L{\Lambda}
\def\F{\Phi}
\def\P{\Psi}

\def\S{\Sigma}

\def\o{\over}
\def\beq{\begin{eqnarray}}
\def\eeq{\end{eqnarray}}
\newcommand{\gsim}{ \mathop{}_{\textstyle \sim}^{\textstyle >} }
\newcommand{\lsim}{ \mathop{}_{\textstyle \sim}^{\textstyle <} }
\newcommand{\vev}[1]{ \left\langle {#1} \right\rangle }
\newcommand{\bra}[1]{ \langle {#1} | }
\newcommand{\ket}[1]{ | {#1} \rangle }
\newcommand{\EV}{ {\rm eV} }
\newcommand{\KEV}{ {\rm keV} }
\newcommand{\MEV}{ {\rm MeV} }
\newcommand{\GEV}{ {\rm GeV} }
\newcommand{\TEV}{ {\rm TeV} }
\def\diag{\mathop{\rm diag}\nolimits}
\def\Spin{\mathop{\rm Spin}}
\def\SO{\mathop{\rm SO}}
\def\O{\mathop{\rm O}}
\def\SU{\mathop{\rm SU}}
\def\U{\mathop{\rm U}}
\def\Sp{\mathop{\rm Sp}}
\def\SL{\mathop{\rm SL}}
\def\tr{\mathop{\rm tr}}
\def\cnb{C$\nu$B~}

\def\IJMP{Int.~J.~Mod.~Phys. }
\def\MPL{Mod.~Phys.~Lett. }
\def\NP{Nucl.~Phys. }
\def\PL{Phys.~Lett. }
\def\PR{Phys.~Rev. }
\def\PRL{Phys.~Rev.~Lett. }
\def\PTP{Prog.~Theor.~Phys. }
\def\ZP{Z.~Phys. }


\title{Coherent Propagation of PeV Neutrinos and the Dip in the Neutrino Spectrum at IceCube}

\author{Ayuki Kamada}
 \email{ayuki.kamada@ucr.edu}
\author{Hai-Bo Yu}%
 \email{haiboyu@ucr.edu}
 \affiliation{%
 Department of Physics and Astronomy, University of California, Riverside, CA 92521, USA
}%

\date{\today}

\begin{abstract}

The energy spectrum of high-energy neutrinos reported by the IceCube collaboration shows a dip between 400 TeV and 1 PeV. One intriguing explanation is that high-energy neutrinos scatter with the cosmic neutrino background through a $\sim$ MeV mediator. Taking the density matrix approach, we develop a formalism to study the propagation of PeV neutrinos in the presence of the new neutrino interaction. If the interaction is flavored such as the gauged $L_\mu-L_\tau$ model we consider, the resonant collision may not suppress the PeV neutrino flux completely. The new force mediator may also contribute to the number of effectively massless degrees of freedom in the early universe and change the diffusion time of neutrinos from the supernova core. Astrophysical observations such as Big Bang Nucleosynthesis and supernova cooling provide an interesting test for the explanation.

\end{abstract}

\maketitle

\section{Introduction}
\label{sec: introduction}
The IceCube experiment has recently reported the observation of neutrinos in the energy range of TeV-PeV\,\cite{Aartsen:2013bka, Aartsen:2013jdh, Aartsen:2014gkd, Aartsen:2014muf}, which provides the first evidence for extraterrestrial high-energy neutrinos. An interesting feature of the observed spectrum is a null detection of high-energy neutrinos in the energy range of $400$-$800 \, {\rm TeV}$. Although at present statistics have not been sufficient enough to confirm the existence of the dip in the spectrum, there have been investigations whether it can be explained by some new physics\,\cite{Esmaili:2013gha,Bai:2013nga,Ioka:2014kca,Zavala:2014dla,Ng:2014pca, Bhattacharya:2014yha, Ibe:2014pja, Blum:2014ewa, Araki:2014ona, Esmaili:2014rma,Cherry:2014xra,Chen:2014gxa}. One possibility is that the high-energy neutrinos may scatter with the cosmic neutrino background (C$\nu$B) and lose their energy, resulting in the dip-like feature in the spectrum\,\cite{Ng:2014pca, Ibe:2014pja, Blum:2014ewa, Araki:2014ona, Cherry:2014xra}.

This scenario has several interesting implications. To suppress the neutrino flux in the range of $400$-$800\, {\rm TeV}$ as indicated by the IceCube observation, the scattering cross section between the high-energy and \cnb neutrinos must be significantly large, which can be achieved by the Breit-Wigner resonance. Since the resonance mass $m_R$ is close to the center-of-mass energy, it can be estimated as $m_R\simeq\sqrt{2m_\nu E_\nu}\sim {\cal O}(1)~{\rm MeV}$, where neutrino energy $E_\nu\sim{\rm PeV}$ and neutrino mass $m_\nu\sim0.01~{\rm eV}$. Therefore, it predicts a new interaction in the neutrino sector with a force mediator much lighter than the weak scale. Because the resonant cross section is sensitive to $E_\nu$ and $m_\nu$, the IceCube neutrino spectrum may contain rich information about neutrino mass and redshift of the source\,\cite{Ibe:2014pja, Blum:2014ewa}.

From the perspective of particle physics model building, it is quite challenging to extend the lepton sector of the Standard Model (SM) with an additional interaction. For example, if the light mediator couples to the three generations of leptons universally, there are strong constraints on the interaction strength from such as electron beam-dump experiments\,\cite{Bjorken:2009mm} and rare decays of mesons \,\cite{Lessa:2007up}. Therefore, it is reasonable to consider models in which the new interaction is not flavor-blind. In this case, it is important to treat the propagation of the high-energy neutrinos properly, in order to calculate the neutrino flux at IceCube. In this paper, we use the density matrix approach to study propagation of PeV neutrinos from the source to the IceCube detector in the presence of a new flavored neutrino self-interaction. To illustrate our main point, we consider an extension of the SM with a gauged $L_{\mu}-L_{\tau}$\,\cite{Baek:2001kca, Ma:2001md, Ma:2002df, Choubey:2004hn, Adhikary:2006rf, Ota:2006xr, Heeck:2010pg, Heeck:2011wj, Baek:2015mna}. This model has several attractive features: it is gauge anomaly-free; it explains the nearly maximum mixing angle between the second and third generations;  the model also evades severe constraints from electron beam-dump experiments. 

We also study cosmological and astrophysical implications of the model. Since the mediator mass is close to the temperature of the Big Bang Nucleosynthesis (BBN) era, the presence of the mediator in the early universe can potentially contribute to the number of effectively massless degrees of freedom. In additional, the light mediator may also be produced in the core of supernovae. The frequent collision between neutrinos mediated by the new force may significantly reduce the neutrino mean free path, which slows down the supernova cooling process. We show that both BBN and supernova constraints are sensitive to the parameter region of the model explaining the dip of the IceCube PeV neutrino spectrum. Our result can be generalized to other models with an ${\cal O}(1)~{\rm MeV}$ force carrier coupled to SM neutrinos.

The rest of the paper is organized as follows. In the next section, we discuss the generic feature of gauged $L_{\mu}-L_{\tau}$ model and experimental constraints. In Sec.\,\ref{sec:neuprop}, we derive the Boltzmann equation governing the evolution of the neutrino density matrix. Then, we discuss cosmological and astrophysical implications in Sec.\,\ref{sec:candaconstrints}. Sec.\,\ref{sec:summary} is devoted to summarizing our results. In APPENDIX, we present the derivation of the resonant scattering rate.

\section{Particle Physics Model}
\label{sec:lmu-ltaumodel}
We assume SM leptons have a new interaction with the following Lagrangian
\begin{eqnarray}
\label{eq:lagrangian}
{\mathcal L}^{\prime}_{\rm int}=- g^{\prime} Z^{\prime}_{\mu} \sum_{\ell, \ell^{\prime}} \left[  {\bar L}_{\ell} \gamma^{\mu} {\mathscr Q}_{\ell \ell^{\prime}} L_{\ell^{\prime}} + {\bar \ell}_{R} \gamma^{\mu} {\mathscr Q}_{\ell \ell^{\prime}} \ell^{\prime}_{R} \right] \,,
\end{eqnarray}
where $g'$ is the coupling constant, $Z'_\mu$ is the new gauge boson, $L_\ell$ denotes the lepton doublet, $\ell_R$ denotes the lepton singlet, and the charge matrix is ${\mathscr Q}_{\ell \ell^{\prime}} = \diag( 0, 1, -1)$
in the interaction basis $\ell = (e, \mu, \tau)$. We furthermore assume that the gauge boson mass is $m_{Z'}\sim{\rm MeV}$.

There are several experimental constraints on this model. The existence of the light $Z'_\mu$ opens up new decay channels for $W$ and $Z$ bosons such as three-body decays $W^{+} \to \mu^{+} \nu Z^{\prime}$. These new processes change $W/Z$-boson decay branching ratios by ${\mit \Delta} \Gamma / \Gamma = 3g^{\prime \, 2}/16\pi^{2}$ in the limit of $m_{Z^{\prime}} \ll m_{W/Z}$\,\cite{Carone:1994aa}.
To achieve $\sim 1 \%$ precision of measured $W/Z$-boson decay branching ratios\,\cite{Agashe:2014kda},
we estimate $g^{\prime} < 0.7$.

One of the most stringent constraints on $g'$ is from the precise measurement of muon anomalous magnetic moment $a_{\mu}=(g-2)/2$. 
The leading contribution from $Z^{\prime}$-exchange to $a_\mu$ can be evaluated as\,\cite{Baek:2001kca, Lynch:2001zr, Heeck:2011wj}
\begin{eqnarray}
{\mit \Delta} a^{Z'}_{\mu} = \frac{g^{\prime \, 2}}{8\pi^{2}} 
\int^{1}_{0} dx \frac{2 m_{\mu}^{2} x^{2} (1-x)}{x^{2} m_{\mu}^{2} + (1-x) m_{Z^{\prime}}^{2}} \,.
\end{eqnarray}
In fact, several experiments have reported the measured value deviates from the SM prediction at the level of ${\mit \Delta} a^{\rm exp}_{\mu} = (42.6 \pm 16.5) \times 10^{-10}$\,\cite{Davier:1998si, Czarnecki:2001pv}. While the hadronic uncertainty is still in debate, we require ${\mit \Delta} a^{Z'}_{\mu}$ to be less than ${\mit \Delta} a^{\rm exp}_{\mu} $, which gives rise to $g^{\prime} \lesssim 5\times10^{-4}$.

The measurements of neutrino-electron interactions also put stringent constraints on the model. Although the $Z'$ boson does not couple to electrons directly in our model, it contributes to neutrino-electron scattering through photon-$Z^{\prime}$ mixing radiatively induced by vacuum polarization with $\mu$ and $\tau$ in the loop. Ref.\,\cite{Harnik:2012ni} analyzed data from Borexino (solar neutrino)\,\cite{Bellini:2011rx} and GEMMA (reactor neutrino)\,\cite{Beda:2009kx} experiments to put constraint on the gauge coupling constant $g_{\rm B-L}$ of a gauged $B-L$ model, which is flavor-blind. To apply their result to our model, we first relax Borexino constraint on $g_{\rm B-L}$ by a factor of $(1/0.66)^{1/4}$, because the $Z'$ boson couples only to $\nu_\mu$ and $\nu_\tau$, but not $\nu_{e}$, which accounts for about $34\,\%$ of the total solar neutrino flux\,\cite{Agashe:2014kda}. We then impose the scaled upper bound on $\sqrt{\epsilon e g'}$, where $e$ is the electric charge of the electron, and $\epsilon$ is the photon-$Z'$ mixing parameter. We calculate the mixing parameter $\epsilon$ as 
\begin{eqnarray}
\epsilon = - \frac{e g'}{2\pi^{2}} \int^{1}_{0} dx \, x(1-x) \ln \left[ \frac{m_{\tau}^{2}-x(1-x)q^{2}}{m_{\mu}^{2}-x(1-x)q^{2}} \right] \,,
\end{eqnarray}
where $m_{\mu}$ and $m_{\tau}$ are the masses of $\mu$ and $\tau$, respectively. We take a typical value of momentum transfer in neutrino-electron scattering, $q^{2}=-1\,{\rm MeV}^{2}$. Note the choice of $q^{2}$ does not change $\epsilon$ as long as $|q^{2}| \ll m_{\mu}^{2}$. We find that constraint from Borexino experiment is more stringent than that from muon anomalous magnetic moment in the light $Z'$ boson region of $m_{Z'}\lesssim10\,{\rm MeV}$. Since the GEMMA experiment looks for reactor ${\bar \nu}_{e}$'s before they oscillate (the distance from reactor is $13.9\,{\rm m}$), it is not applicable to our model because $Z'$ does not couple to $\nu_e$. 

The realization of the observed neutrino masses and mixing angles in the gauged $L_\nu-L_\tau$ model has been discussed in the literature\,\cite{Choubey:2004hn, Adhikary:2006rf, Ota:2006xr, Heeck:2010pg, Heeck:2011wj, Baek:2015mna}. In this paper, we assume that neutrino masses are quasi-degenerate, which can be achieved with a proper choice of the symmetry breaking pattern\,\cite{Heeck:2011wj}. In this case, we can translate the cosmological limit $\sum_i m_{\nu_i} < 0.25\,{\rm eV}$\,\cite{Hinshaw:2012aka, Ade:2013zuv, Aubourg:2014yra} to an upper bound on the individual neutrino mass $m_{\nu_i}<0.083~{\rm eV}$. On the other hand, the observed atmospherical neutrino mass is $\sqrt{{\mit \Delta} m^{2}_{\rm atm}} \simeq 0.048 \,{\rm eV}$, which leads to a lower bound on $m_{\nu_i}\sim0.05~{\rm eV}$ for a degenerate neutrino mass spectrum. Since the dip of the PeV neutrino spectrum at IceCube is in the energy range of $4 \times 10^{2} \lesssim E_{\nu} \lesssim 8 \times 10^{2}\,{\rm TeV}$\,\cite{Aartsen:2014gkd, Aartsen:2014muf} and the resonance condition for neutrino scattering is $m^{2}_{Z^{\prime}}\simeq2 E_{\nu} m_{\nu_i}$, we obtain a preferred range of  ${Z^{\prime}}$ mass $5 \, {\rm MeV} \lesssim m_{Z'} \lesssim 10\, {\rm MeV}$. If scattering occurs at a high redshift, we can shift this mass range by a factor of $\sqrt{1+z}$ accordingly.

\section{Neutrino Propagation}
\label{sec:neuprop}

We first estimate the coherence length of PeV neutrinos as follows~\cite{Fukugita:2003en,Farzan:2008eg,Akhmedov:2012uu}. For any two of the neutrino mass eigenstates composing a flavor state, the velocity difference of their wave packets is $|v_i-v_j|\simeq |\Delta m^2_{ij}|/2E^2_\nu$, where ${\mit \Delta} m^{2}_{ij}\equiv m^{2}_{i} - m^{2}_{j}$. After they travel distance $L$, the wave packets are $\sim L|\Delta m^2_{ij}|/2E^2_\nu$ apart. If $L |\Delta m^2_{ij}|/2E^2_\nu$ is larger than the uncertainty in their spatial location, the wave packets do not overlap and lose coherence.\footnote{In this case, the density matrix defined in Eq.~(\ref{eq:densitymatrix}) is diagonal in the mass basis.} Therefore, the coherence length can be estimated as~\cite{Farzan:2008eg}
\begin{eqnarray}
L_{{\rm coh}, ij}\simeq\frac{4\pi E^2_\nu}{|\Delta m^2_{ij}|}\sigma_x,
\end{eqnarray}
where  $\sigma_x$ is the spatial uncertainty of PeV neutrinos. Assuming that IceCube PeV neutrinos are produced by decays of high-energy pions, we expect the spatial uncertainty of PeV neutrinos is of the order of the distance that the pion travels before it decays, {\it i.e.}, $\sigma_x\simeq m_\pi\tau_\pi/(4E_\nu)$~\cite{Farzan:2008eg}, where $m_\pi/(4E_\nu)$ is the Lorentz contraction factor, and the pion lifetime and mass in the rest frame are $\tau_{\pi} \simeq 2.6\times 10^{-8}\,{\rm s}$, and $m_{\pi} \simeq 140\,{\rm MeV}$, respectively. Taking $E_\nu$=1 PeV and $|\Delta m^2_{ij}|=10^{-3}$ eV$^2$, we can estimate the spatial uncertainty $\sigma_x\simeq2.7\times10^{-5}$ cm and the coherence length $L_{{\rm coh}, ij}\simeq 100$ Gpc. Therefore, if PeV neutrinos are produced by decay of free pions, the coherent length can be larger than the particle horizon size of the universe $\sim14~{\rm Gpc}$. However, the environmental effects of the PeV neutrino source, such as collisions of the parent particle with particles in medium and the presence of magnetic fields, may shorten the coherence length significantly~\cite{Farzan:2008eg,Jones:2014sfa,Akhmedov:2014ssa}. Furthermore, even if the coherent oscillation is maintained during the propagation, it may not be detected due to uncertainties of source distance and limitations of detector resolution. In our analysis, we first assume that PeV neutrinos are coherent and derive the probability matrix in the presence of the new interaction, and then take time-averaging over the oscillatory terms to include the possible decoherence effects. Therefore, the formalism we will develop below is valid even if PeV neutrinos have a coherent length shorter than the propagation distance.


To describe propagation of the PeV neutrinos from the source to the IceCube detector, we consider evolution of the following density matrix
\begin{eqnarray}
{\mathscr F}_{\ell \ell^{\prime}}({\vec k}, t) \propto \langle a^{\dagger}_{\ell}({\vec k}) a_{\ell^{\prime}}({\vec k}) \rangle \,,
\label{eq:densitymatrix}
\end{eqnarray}
where $a_{\ell}({\vec k})$ is an annihilation operator of lepton $\nu_\ell$ neutrino with momentum $\vec k$. We normalize the density matrix such that the number density is given by $n_{\ell} (t)= \int d^{3}{\vec k}/(2\pi)^{3} {\mathscr F}_{\ell \ell}({\vec k}, t)$. The evolution equation of the density matrix can be derived from nonequilibrium field theory\,\cite{Cirigliano:2009yt, Cirigliano:2011di, Tulin:2012re},
\begin{eqnarray}
\frac{\partial}{\partial t} {\mathscr F}({\vec k}, t) 
- H {\vec k} \frac{\partial}{\partial {\vec k}} {\mathscr F}({\vec k}, t) 
= 
- i [{\mathscr H}(\vec k), {\mathscr F}(\vec k, t) ]
+{\mathscr C}[{\mathscr F}] \,,
\end{eqnarray}
where $H$ is the Hubble expansion rate, $\left[ \cdot, \cdot \right]$ denotes the commutator, ${\mathscr H}(\vec k)$ is the Hamiltonian, and ${\mathscr C}[{\mathscr F}]$ represents the collision term. The Hamiltonian is
\begin{eqnarray}
{\mathscr H}(\vec k) = \sqrt{k^{2} + {\mathscr M}_{\nu}^{*}{\mathscr M}_{\nu}} \simeq \omega(k) + {\mit \Delta}{\mathscr M}_{\nu}^{*}{\mathscr M}_{\nu}/ (2\omega(k)) \,,
\end{eqnarray}
where $k=|{\vec k}|$, $\omega(k)=\sqrt{k^{2}+m_{\nu}^{2}}$, $m_{\nu}^{2} = {\rm tr} ({\mathscr M}_{\nu}^{*}{\mathscr M}_{\nu}) / 3$, and ${\mit \Delta}{\mathscr M}_{\nu}^{*}{\mathscr M}_{\nu} = {\mathscr M}_{\nu}^{*}{\mathscr M}_{\nu}- m_{\nu}^{2}$. Since mass-squared differences are small, we have ${\mit \Delta}m^{2} \ll \omega(k)$. 

We write the collision term as
\begin{eqnarray}
{\mathscr C}[{\mathscr F}] 
=
-\frac{\Gamma_{s}({\vec k}, t)}{2} \{ {\mathscr F}({\vec k}, t), {\mathscr R} \}  \,,
\end{eqnarray}
and the total scattering rate is given by
\begin{eqnarray}
\label{eq:totalrate}
\Gamma_{s}({\vec k}, t)
= 9 \zeta(3) T_{\nu}^{3}(t) \frac{1}{m_{\nu}} \frac{\Gamma_{Z^{\prime}}}{m_{Z^{\prime}}} \delta\left[k-m_{Z^{\prime}}^{2}/(2m_{\nu})\right] \,,
\end{eqnarray}
where $\zeta(s)$ is Riemann zeta function. We present the derivation of Eq.~(\ref{eq:totalrate}) in APPENDIX.

Taking redshift $z$ and incoming momentum ${\vec k}_{0}$ as time and momentum coordinates instead of cosmic time $t$ and physical momentum $\vec k$, respectively, we obtain the density matrix evolution of PeV neutrino,
\begin{eqnarray}
\label{eq:evolution}
-\frac{\partial}{\partial z} {\widetilde {\mathscr F}}({\vec k}_{0}, z) 
= 
- i [{\mit \Delta}{\widetilde {\mathscr H}}({\vec k}_{0}, z), {\widetilde {\mathscr F}}({\vec k}_{0}, z) ]
-\frac{{\widetilde \Gamma_{s}}({\vec k}_{0}, z)}{2} \{{\widetilde {\mathscr F}}({\vec k}_{0}, z), {\mathscr R} \} \,
\end{eqnarray}
where
\begin{align}
\label{eq:redefine}
{\widetilde {\mathscr F}}
= {\mathscr F}({\vec k}, t),~
{\mit \Delta}{\widetilde {\mathscr H}}
= \frac{{\mit \Delta}{\mathscr M}_{\nu}^{*}{\mathscr M}_{\nu}}{2k_{0}(1+z)^{2}H(z)},~
{\widetilde \Gamma_{s}}
= {\widetilde \tau_{s}}(z) \delta\left[1+z-\frac{m_{Z^{\prime}}^{2}}{2m_{\nu}k_{0}} \right],
\end{align}
and the optical depth ${\widetilde \tau_{s}}(z)$ is
\begin{align}
\label{eq:depth}
{\widetilde \tau_{s}}(z)=18 \zeta(3) T_{\nu, 0}^{3} (1+z)^{3} \frac{1}{H(z)m_{Z^{\prime}}^{2}} \frac{\Gamma_{Z^{\prime}}}{m_{Z^{\prime}}}.
\end{align}

Integrating both sides of Eq.~(\ref{eq:evolution}) from $z_{i}$ to $z_{f}$, we obtain a formal solution for ${\widetilde {\mathscr F}}$
\begin{eqnarray}
{\widetilde {\mathscr F}}({\vec k}_{0}, z_{f}) 
= {\widetilde {\mathscr P}}({\vec k}_{0}, z_{f}, z_{i}) {\widetilde {\mathscr F}}({\vec k}_{0}, z_{i}) {\widetilde {\mathscr P}}({\vec k}_{0}, z_{f}, z_{i})^{\dagger}
\end{eqnarray}
with non-unitary operator
\begin{eqnarray}
\label{eq:timeoperator}
{\widetilde {\mathscr P}}({\vec k}_{0}, z^{\prime}, z) 
= {\cal P} \left\{ \exp\left(-i \int^{z}_{z^{\prime}} dz^{\prime\prime} \left[ {\widetilde {\mathscr H}}({\vec k}_{0}, z^{\prime\prime}) 
-i {\widetilde \Gamma_{s}}({\vec k}_{0}, z^{\prime\prime}) {\mathscr R} / 2 \right] \right)\right\} \,,
\end{eqnarray}
where ${\cal P}$ is the propagation order operator defined such that ${\mathcal P}\left\{ {\cal Q}(z) {\cal Q}^{\prime}(z^{\prime}) \right\} = {\cal Q}(z){\cal Q}^{\prime}(z^{\prime})$ for $z < z^{\prime}$ and ${\cal Q}^{\prime}(z^{\prime}){\cal Q}(z)$ for $z^{\prime} < z$.
Substituting Eq.~(\ref{eq:redefine}) into Eq.~(\ref{eq:timeoperator}), we obtain
\begin{eqnarray}
{\widetilde {\mathscr P}}({\vec k}_{0}, z^{\prime}, z) 
= {\widetilde {\mathscr U}}({\vec k}_{0}, z^{\prime}, z_{s}(k_{0})) {\widetilde {\mathscr T}_{s}}(z_{s}(k_{0}), z^{\prime}, z) {\widetilde {\mathscr U}}({\vec k}_{0}, z_{s}(k_{0}), z) \,,
\end{eqnarray}
where
\begin{eqnarray}
{\widetilde {\mathscr U}}({\vec k}_{0}, z^{\prime}, z) 
&=& \exp \left[-i \frac{{\mit \Delta}{\mathscr M}_{\nu}^{*}{\mathscr M}_{\nu}}{2k_{0}} {\widetilde L}(z^{\prime}, z) \right]~{\rm with}~{\widetilde L}(z^{\prime}, z)=\int^{z}_{z^{\prime}} \frac{dz^{\prime\prime}} {(1+z^{\prime\prime})^{2}H(z^{\prime\prime})} \,, \\
\nonumber{\widetilde {\mathscr T}_{s}}(z^{\prime\prime},z^{\prime},z) 
&=& 
\begin{cases}
\exp\left(- {\widetilde \tau_{s}}(z^{\prime\prime}) {\mathscr R} / 2 \right) & (z^{\prime} < z^{\prime\prime} < z) \\
1 & ({\rm otherwise})
\end{cases} \,, \\
z_{s}(k_{0})&=& m_{Z^{\prime}}^{2} / (2m_{\nu}k_{0})-1\,.
\end{eqnarray}
$U$ is the unitary matrix that relates the mass basis $\nu_{i}$ to a basis of interest $\nu_{\ell} = \sum_{i} U_{\ell i} \nu_{i}$, the density matrix at $z=z_{f}$ can be written as
\begin{eqnarray}
\nonumber{\widetilde {\mathscr F}}_{\ell \ell^{\prime}}(z_{f})
=&&\sum_{\ell'{\rm s}, i'{\rm s}} 
U_{\ell i_{3}} U^{*}_{\ell_{5} i_{3}}  {\widetilde {\mathscr T}_{s, \ell_{5} \ell_{3}}}(z_{s}, z_{f}, z_{i}) U_{\ell_{3} i_{1}} U^{*}_{\ell_{1} i_{1}}
{\widetilde {\mathscr F}}_{\ell_{1} \ell_{2}}(z_{i})
U_{\ell_{2} i_{2}} U^{*}_{\ell_{4} i_{2}} {\widetilde {\mathscr T}_{s, \ell_{6} \ell_{4}}}(z_{s}, z_{f}, z_{i}) U_{\ell_{6} i_{4}} U^{*}_{\ell^{\prime} i_{4}}\\ 
&&\times \exp\left[-i \frac{{\mit \Delta} m^{2}_{i_{1}i_{2}}}{2k_{0}} {\widetilde L}(z_{s}, z_{i}) \right]
\exp\left[-i \frac{{\mit \Delta} m^{2}_{i_{3}i_{4}}}{2k_{0}} {\widetilde L}(z_{f}, z_{s}) \right].
\end{eqnarray}

For PeV neutrinos, the oscillation length $L_{{\rm osc}, ij}=4\pi k_{0}/{\mit \Delta} m^{2}_{ij}$ is $8 \times 10^{-8} (k_{0}/{\rm PeV}) ({\rm eV}^{2}/m^{2}_{ij})\, {\rm pc}$, while the propagation length ${\widetilde L}(z^{\prime}, z)$ is on the order of $1/H_{0} = 3 \times 10^{3} \,{\rm Mpc}/h$ as long as $z^{\prime}$ and $z - z^{\prime}$ are on the order of unity. 
This implies that we can take the period average of the exponential terms, which gives rise to $\left<\exp[-i {\mit \Delta} m^{2}_{ij} {\widetilde L} / (2k_{0})]\right>\simeq \delta_{ij}$.
With this approximation, we obtain the density matrix as
\begin{align}
{\widetilde {\mathscr F}}_{\ell \ell^{\prime}}(z_{f})
=\sum_{\ell'{\rm s}, i'{\rm s}} 
U_{\ell i_{2}} U^{*}_{\ell_{5} i_{2}}  {\widetilde {\mathscr T}_{s, \ell_{5} \ell_{3}}}(z_{s}, z_{f}, z_{i}) U_{\ell_{3} i_{1}} U^{*}_{\ell_{1} i_{1}} {\widetilde {\mathscr F}}_{\ell_{1} \ell_{2}}(z_{i})
U_{\ell_{2} i_{1}} U^{*}_{\ell_{4} i_{1}} {\widetilde {\mathscr T}_{s, \ell_{6} \ell_{4}}}(z_{s}, z_{f}, z_{i}) U_{\ell_{6} i_{2}} U^{*}_{\ell^{\prime} i_{2}}.
\end{align}
For PeV neutrino detection, it is useful to take the interaction basis.
Here, the unitary matrix is called Maki-Nakagawa-Sakata (MNS) matrix\,\cite{Maki:1962mu} that is often parametrized by
\begin{eqnarray}
U_{\rm MNS}
=
\left[
\begin{array}{ccc}
1 & 0 & 0 \\
0 & c_{23} & s_{23} \\
0 & -s_{23} & c_{23}
\end{array}
\right]
\left[
\begin{array}{ccc}
c_{13} & 0 & s_{13} e^{-i \delta} \\
0 & 1 & 0 \\
-s_{13} e^{i \delta} & 0 & c_{13}
\end{array}
\right]
\left[
\begin{array}{ccc}
c_{12} & s_{12} & 0 \\
-s_{12} & c_{12} & 0 \\
0 & 0 & 0
\end{array}
\right]
\diag(1,e^{i \alpha_{21}/2}, e^{i \alpha_{31}/2}) \,,
\end{eqnarray}
with $c_{ij}$ and $s_{ij}$ denote $\cos(\theta_{ij})$ and $\sin(\theta_{ij})$, respectively.
In the interaction basis, the interaction matrix is diagonal,
\begin{eqnarray}
\label{eq:scatteringterm}
{\widetilde {\mathscr T}_{s}}(z^{\prime\prime},z^{\prime},z) 
&=& 
\begin{cases}
\diag(1, e^{- {\widetilde \tau_{s}}(z^{\prime\prime}) / 2}, e^{- {\widetilde \tau_{s}}(z^{\prime\prime}) / 2}) & (z^{\prime} < z^{\prime\prime} < z) \\
1 & ({\rm otherwise})
\end{cases} \,.
\end{eqnarray}
The elements of the probability matrix are
\begin{eqnarray}
{\mathscr P}_{\ell \ell^{\prime}} 
={\widetilde {\mathscr F}}_{\ell \ell}({\vec k}_{0}, z_{f}) \quad {\rm for}  \quad {\widetilde {\mathscr F}}_{\ell_{1} \ell_{2}}({\vec k}_{0}, z_{i}) = \delta_{\ell_{1} \ell^{\prime}} \delta_{\ell_{2} \ell^{\prime}}
|\langle \nu_{\ell}(z_{f})|\nu_{\ell^{\prime}}(z_{i})\rangle|^{2},
\end{eqnarray}
where we do not sum over $\ell'$ and $\ell$. We can write ${\mathscr P}_{\ell \ell^{\prime}}$ explicitly as
\begin{eqnarray}
\label{eq:pmatrix}
{\mathscr P}_{\ell \ell^{\prime}} =&& \sum_{\ell'{\rm s}, i'{\rm s}} 
U_{\ell i_{2}} U^{*}_{\ell_{1} i_{2}}  {\widetilde {\mathscr T}_{s, \ell_{1} \ell_{1}}}(z_{s}, z_{f}, z_{i}) U_{\ell_{1} i_{1}} U^{*}_{\ell^{\prime} i_{1}}U_{\ell^{\prime} i_{1}} U^{*}_{\ell_{2} i_{1}} {\widetilde {\mathscr T}_{s, \ell_{2} \ell_{2}}}(z_{s}, z_{f}, z_{i}) U_{\ell_{2} i_{2}} U^{*}_{\ell i_{2}} \,.
\end{eqnarray}
To evaluate ${\mathscr P}$ numerically, we take the following values for the MNS matrix parameters (normal hierarchy)\,\cite{Agashe:2014kda}:
$\sin^{2}(\theta_{12})=0.308,~\sin^{2}(\theta_{23})=0.437,~\sin^{2}(\theta_{13})=0.0234,$ and $\delta/\pi=1.39$. Combining Eqs.~(\ref{eq:scatteringterm}) and (\ref{eq:pmatrix}), we obtain
\begin{equation}
\label{eq:probability}
{\mathscr P}
\simeq 
\left[
\begin{array}{ccc}
0.30 & 0.13 & 0.12 \\
0.13 & 0.06 & 0.05 \\
0.12 & 0.05 & 0.04
\end{array}
\right]
+ e^{- {\widetilde \tau_{s}}(z_s) / 2}
\left[
\begin{array}{ccc}
0.07 & -0.05 & - 0.03 \\
-0.05 & 0.03 & 0.02 \\
- 0.03 & 0.02 & 0.01
\end{array}
\right]
+ e^{- {\widetilde \tau_{s}}(z_s)}
\left[
\begin{array}{ccc}
0.18 & 0.15 & 0.12 \\
0.15 & 0.29 & 0.31 \\
0.12 & 0.31 & 0.35
\end{array}
\right],
\end{equation}
for $z_s\geq0$.

With the probability matrix given in Eq.~(\ref{eq:probability}), we check several extreme cases. In the absence of scattering, {\it i.e.,} $ {\widetilde \tau_{s}}(z_s)=0$, the flavor composition of PeV neutrinos at the IceCube detector is completely determined by the initial condition and oscillations. From Eq.~(\ref{eq:probability}), we can see that $\sum_{\ell'(\ell)}{\mathscr P}_{\ell\ell'}=1$, where $\ell'(\ell)=e,\mu,\tau$. This is expected because without scattering, the total probability for finding neutrinos in different flavors is conserved.  

In the limit of ${\widetilde \tau_{s}}(z_s)\gg1$, one might think that $\nu_\mu$ and $\nu_\tau$ would be completely depleted in the neutrino flux reaching the IceCube detector because of collisions mediated by the $Z'$. However, this is not the case. Even though the last two terms of Eq.~(\ref{eq:probability}) vanish, the first one does not depend on ${\widetilde \tau_{s}}(z_s)$ at all. Therefore, the probability for finding $\nu_\mu$ and $\nu_\tau$ does {\it not} vanish. In general, for a model with flavor-dependent neutrino interactions, the \cnb can not completely absorb each flavor of high-energy neutrinos.

\begin{figure}[t]
\begin{center}
\includegraphics[width=0.9\linewidth]{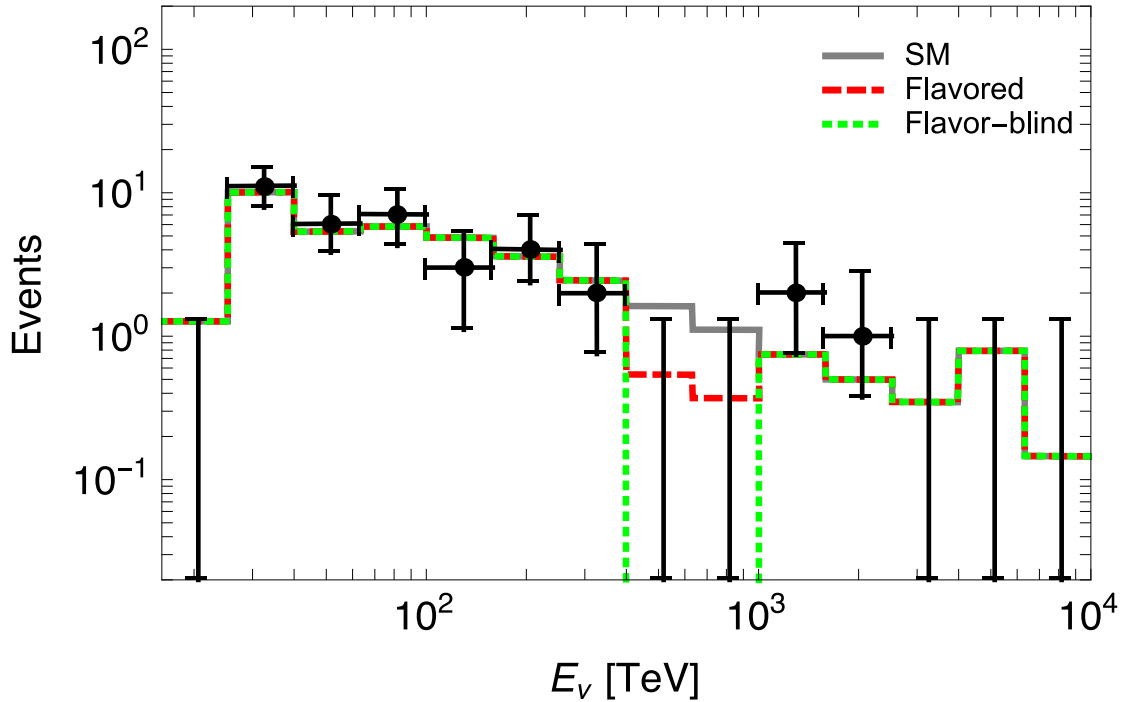}
\caption{\sl \small
Expected number of PeV neutrino events at the IceCube detector for the standard model (solid), the flavored $L_{\mu}-L_{\tau}$ model (dotted), and a model in which neutrinos have a new flavor-blind interaction (dashed), together with experimental data~\cite{Aartsen:2014gkd}. For the $L_{\mu}-L_{\tau}$ model, we assume that PeV neutrinos originate from $pp$ collisions. Because of the coherent effect in PeV neutrino propagation, the neutrino flux does not vanish near the resonance for the flavored model. 
}
\label{fig:neutrinoevent}
\end{center}
\end{figure}

To check whether the $L_\mu-L_\tau$ model can produce the dip in the energy spectrum of high-energy neutrinos observed by the IceCube, we consider two possible sources for the high-energy neutrinos. If they originate from $pp$ collisions, the initial flavor compositions in the neutrino flux are $(\phi_{\nu_{e}}, \phi_{\nu_{\mu}}, \phi_{\nu_{\tau}}) = (\phi_{{\bar \nu}_{e}}, \phi_{{\bar \nu}_{\mu}}, \phi_{{\bar \nu}_{\tau})} \simeq (1, 2, 0)$. Applying the probability matrix of Eq.~(\ref{eq:probability}) with ${\widetilde \tau_{s}}(z_s)\gg1$, we obtain the final flavor composition is $(0.56, 0.25, 0.22)$, and the total survival rate is $1/3$. While, for a $p\gamma$ source, the initial flavor compositions are $(1, 1, 0) $ and $(0, 1, 0)$ for neutrinos and anti-neutrino, respectively. At the detector, they become $(0.43, 0.19, 0.17)$ with a $2/5$ survival rate and $(0.13, 0.06, 0.05)$ with a $1/4$ survival rate, respectively. Such a fractional suppression can explain the dip in the high-energy neutrino spectrum observed in the IceCube detector (see Fig.\,\ref{fig:neutrinoevent}). Within the current data set, the expected number of events is just $2$. Therefore, even a $2/5$ suppression factor can reduce the expected number of events below the Poisson limit, $1$. On the other hand, our result can be tested with better statistics in the accumulated data set in the near future.

To examine the favored parameter region for the $L_\mu-L_\tau$ model, we use Eq.~(\ref{eq:depth}) and take $m_{Z'}=8~{\rm MeV}$, $m_\nu=0.05~{\rm eV}$, and $k_0=600~{\rm TeV}$. The Hubble expansion rate is $H(z)=H_0\sqrt{\Omega_M(1+z)^3+\Omega_\Lambda}$, where $H_0=100h~{\rm km/s/Mpc}$, $\Omega_M=0.32$, $\Omega_\Lambda=0.68$, and $h=0.69$. We find the redshift at which the collision occurs is $z_s\simeq0.07$, and the optical depth is
\begin{equation}
{\widetilde \tau_{s}}\simeq1\left(\frac{g'}{1.7\times10^{-4}}\right)^2.
\end{equation}
We see that the required value $g'$ for the model to explain the dip is below the constraint from the muon $g-2$ measurement, $g'\lesssim5\times10^{-4}$. It is interesting to note that the model can explain the dip in the neutrino spectrum and the discrepancy in the muon anomalous magnetic moment. If we take $g'=5\times10^{-4}$ as preferred by ${\mit \Delta} a^{\rm exp}_{\mu}$, the optical depth is ${\widetilde \tau_{s}}\simeq8.5$, which is more than enough to suppress the neutrino flux in the 400-800 TeV energy range at the IceCube detector~\cite{Araki:2014ona}. However, as we will show later, this parameter region is strongly disfavored by the constraint from supernova cooling.  

Our result can be easily generalized to the case in which the neutrino interaction is flavor-blind. After replacing ${\mathscr R}={\rm diag}(0,1,1)$ by ${\rm diag}(1,1,1)$, we can write the probability matrix as 
\begin{equation}
\label{eq:probability2}
{\mathscr P}
\simeq 
e^{- {\widetilde \tau_{s}}(z_s)}
\left[
\begin{array}{ccc}
0.54 & 0.24 & 0.21 \\
0.24 & 0.38 & 0.38 \\
0.21 & 0.38 & 0.41
\end{array}
\right].
\end{equation}
In this case, all flavor compositions in the neutrino flux will be suppressed if $\tau_s(z_s)\gg1$.

\section{Cosmological and Astrophysical implications}
\label{sec:candaconstrints}
In this section, we study cosmological and astrophysical implications if neutrinos interact with a new $\sim$ MeV force carrier. We take the $L_{\mu}-L_{\tau}$ model as an example, and our analysis can be generalized to other models. 
\subsection{$\Delta N_{\rm eff}$ constraints}
\label{subsec:deltaNeff}

In the early universe, $Z^{\prime}$ bosons can be produced in the SM thermal both by inverse decay and pair annihilation of leptons. The rate of inverse decay can be estimated as $\Gamma_{\rm inv}\sim g^{\prime 2} m_{Z^{\prime}}\times m_{Z^{\prime}}/T$, where $m_{Z^{\prime}}/T$ is the time dilation factor. The rate for pair annihilation is $\Gamma_{\rm ann}\sim {g'}^4T$. At early stages when the temperature is high, the pair annihilation process dominates the production of $Z'$ in the thermal bath. But inverse decay becomes more important when $T \lesssim 100\,{\rm GeV} (10^{-4}/g^{\prime}) (m_{Z^{\prime}}/10 \, {\rm MeV})$. When the temperature drops below $m_{Z^{\prime}}$, the number density of $Z'$ becomes suppressed by the Boltzmann factor. Since the $Z'$ mass is close to the temperature when BBN starts, it may contribute to the effective number of neutrinos $N_{\rm eff}$. 
 
The mediator $Z'$ may change $N_{\rm eff}$ in two ways. If $Z'$ is still relativistic during BBN, it contributes to $N_{\rm eff}$ directly. In this case, ${\mit \Delta} N_{\rm eff} = 3\times4/7 \simeq 1.7$, which is strongly disfavored by observations of light nuclei abundances\,\cite{Mangano:2011ar, Steigman:2012ve} and CMB anisotropies~\cite{Hinshaw:2012aka, Ade:2013zuv, Aubourg:2014yra}. Since $m_{Z'}\sim$ 10 MeV in the model we consider, the direct contribution to $N_{\rm eff}$ at $T$=0.1-1 MeV is negligible because of the Boltzmann suppression factor. However, even in this case, $Z'$ may still contribution to $N_{\rm eff}$ in an indirect way. When $Z'$ becomes nonrelativistic, it transfers its entropy to $\nu_\mu$ and $\nu_\tau$ and increase their temperature relative to the temperature of $\nu_e$ after neutrinos decouple from the SM thermal bath at $T_{\nu,{\rm dec}}$=1.5 MeV. To study this subtle effect, we take the following steps. We assume all neutrinos and anti-neutrinos have the same temperature $T_{\nu,{\rm dec}}$=1.5 MeV when they decouple from the SM thermal bath. After decoupling, the $\nu_\mu$ and $\nu_\tau$, and $Z'$ form a thermal bath, which evolve independently from $\nu_e$ and the photon. Then, we follow the phase space distribution function of $\nu_\mu$, $\nu_\tau$, and $Z'$ from $T_{\nu,{\rm dec}}$=1.5 MeV to $T_\gamma=$0.1 MeV, and derive a lower bound on $m_{Z'}$ by demanding ${\mit \Delta} N_{\rm eff}<0.7$ at $T=$0.1 MeV.
      
Since both inverse decay and pair annihilation processes respect $CP$, the relevant phase space distribution functions (per spin degrees of freedom) are given by
\begin{eqnarray}
\label{eq:distribution}
f_{\nu_{\mu}} = f_{{\bar \nu}_{\mu}} = f_{\nu_{\tau}} = f_{{\bar \nu}_{\tau}} = \frac{1}{e^{k/T' - \xi}+1},~f_{Z^{\prime}} = \frac{1}{e^{\sqrt{k^{2}+m_{Z^{\prime}}^{2}}/T' - \xi^{\prime}} -1} \,.
\end{eqnarray}
where $T'$ denotes the temperature of $\nu_\nu$, $\nu_\tau$ and $Z'$ after they decouple from the SM thermal bath, $\xi$ and $\xi'$ are the chemical potential per unit temperature for the neutrinos and $Z'$ bosons, respectively. 

To evaluate $\xi$ and $\xi^{\prime}$, we impose the following three conditions.
 \begin {itemize}
 \item{$\xi'=2\xi$, because the inverse decay process is in the thermal equilibrium.}
\item{The entropy per comoving volume is conserved, 
\begin{eqnarray}
\label{eq:entropy}
(s_{\nu_{\mu}} + s_{{\bar \nu}_{\mu}} + s_{\nu_{\tau}}  + s_{{\bar \nu}_{\tau}} + s_{Z^{\prime}} ) a^{3}= {\rm constant},
\end{eqnarray}
where $a$ is the scale factor.}
\item{The third condition depends on whether the pair annihilation process is in chemical equilibrium when the number density of $Z'$ becomes negligible. We estimate the equilibrium condition requires $g'\gtrsim 10^{-5}$. In this case, $\xi'=\x$. Combining with the first condition, we have $\xi'=\xi=0$. If not, we instead use the conservation condition of the comoving number density 
\begin{eqnarray}
\label{eq:density}
(n_{\nu_{\mu}} + n_{{\bar \nu}_{\mu}} + n_{\nu_{\tau}}  + n_{{\bar \nu}_{\tau}} + 2 n_{Z^{\prime}} ) a^{3}= {\rm constant}. 
\end{eqnarray}
We will discuss both cases.}
\end{itemize} 

With the distribution functions given in Eq.~(\ref{eq:distribution}), we can write the entropy densities as 
\begin{eqnarray}
s_{\nu_{\mu}} &=& s_{{\bar \nu}_{\mu}} = s_{\nu_{\tau}} = s_{{\bar \nu}_{\tau}} = \int \frac{4\pi k^{2} dk}{(2\pi)^{3}} \left[ \frac{4k}{3T'} - \xi \right] f_{\nu}, \\s_{Z^{\prime}} &=& 3 \int \frac{4\pi k^{2} dk}{(2\pi)^{3}} \left[ \frac{\sqrt{k^{2}+m_{Z'}^{2}}}{T'} + \frac{k^{2}}{3 T' \sqrt{k^{2}+m_{Z'}^{2}}} - \xi^{\prime} \right] f_{Z'}.
\end{eqnarray}

In the case of $\xi'=\xi=0$, we use the entropy conservation condition Eq.~(\ref{eq:entropy}) to determine the temperature, $T'$, at $T$=0.1 MeV for a given $m_{Z'}$. As we know, in the standard case, all neutrino species have the same temperature $0.1\times(4/11)^{1/3}$ MeV when $T$=0.1 MeV. In our case, the $\nu_e$ evolves as before, but both $\nu_\mu$  and $\nu_\tau$ should have a higher temperature than $0.1\times(4/11)^{1/3}$ because they inherit the energy density of the $Z'$ boson. To evaluate the energy densities of the $\nu_\mu$, $\nu_\tau$ and $Z'$, we use
\begin{eqnarray}
\rho_{\nu_{\mu}} = \rho_{{\bar \nu}_{\mu}} = \rho_{\nu_{\tau}} = \rho_{{\bar \nu}_{\tau}} = \int \frac{4\pi k^{2} dk}{(2\pi)^{3}} k f_{\nu},
\rho_{Z^{\prime}} = 3 \int \frac{4\pi k^{2} dk}{(2\pi)^{3}} \sqrt{k^{2}+m_{Z'}^{2}} f_{Z'} \,.
\end{eqnarray}
We find that $\rho_{Z^{\prime}}$ is negligible $m_{Z'}\gtrsim$1 MeV at $T$=0.1 MeV. Using the standard definition,
\begin{eqnarray}
\rho_{\gamma} + \rho_{\nu_{e}} + \rho_{{\bar \nu}_{e}} + \rho_{\nu_{\mu}} + \rho_{{\bar \nu}_{\mu}} + \rho_{\nu_{\tau}} + \rho_{{\bar \nu}_{\tau}} = \rho_{\gamma} \left[ 1+\frac{7}{8} \left(\frac{4}{11}\right)^{4/3} \left( {\mit \Delta} N_{\rm eff}+3\right) \right] \,,
\label{eq:deltan}
\end{eqnarray}
we calculate ${\mit \Delta} N_{\rm eff}$.
Since the presence of the $Z'$ boson does not change the thermal history of $\gamma$ and $\nu_{e}$, the following relation is still valid in our model
\begin{eqnarray}
\rho_{\gamma} + \rho_{\nu_{e}} + \rho_{{\bar \nu}_{e}} = \rho_{\gamma} \left[ 1+ \frac{7}{8} \left(\frac{4}{11}\right)^{4/3} \right].
\end{eqnarray}
We demand ${\mit \Delta} N_{\rm eff}$ defined in Eq.~(\ref{eq:deltan}) to be less than $0.7$, and derive an upper bound $m_{Z^{\prime}} \gtrsim 5.3 \,{\rm MeV}(T_{\nu, {\rm dec}} / 1.5 {\rm MeV})$ shown in Fig.~\ref{fig:summary}. We note that ${\mit \Delta} N_{\rm eff}$ drops significantly for larger $m_{Z^{\prime}}$. For example, ${\mit \Delta} N_{\rm eff}$ is $0.1$ for $m_{Z^{\prime}} \gtrsim 10 \,{\rm MeV}(T_{\nu, {\rm dec}} / 1.5 {\rm MeV})$. This is because the energy density carried by the $Z'$ boson is suppressed by the Boltzmann factor $\sim\exp(-m_{Z'}/T')$.

If the process of pair annihilation and creation becomes less than the Hubble expansion rate when the $Z'$ boson becomes nonrelativistic, $T'$, $\xi$ and $\xi'$ can be determined by solving Eqs.~(\ref{eq:entropy}) and~(\ref{eq:density}) simultaneously with the initial condition $\xi=\xi'=0$ at $T_{\nu,{\rm dec}}=1.5$ MeV, where the number densities are given by
\begin{eqnarray}
\label{eq:nfornu}
n_{\nu_{\mu}} = n_{{\bar \nu}_{\mu}} = n_{\nu_{\tau}} = n_{{\bar \nu}_{\tau}} = \int \frac{4\pi k^{2} dk}{(2\pi)^{3}} f_{\nu}, n_{Z^{\prime}} = 3 \int \frac{4\pi k^{2} dk}{(2\pi)^{3}} f_{Z'} \,.
\label{eq:nforZprime}
\end{eqnarray}

Following a similar procedure, we obtain $m_{Z^{\prime}} \gtrsim 5.3 \,{\rm MeV} (T_{\nu, {\rm dec}} / 1.5 {\rm MeV})$ for ${\mit \Delta} N_{\rm eff}  < 0.7$, which is similar to the upper bound for the case of $\xi'=\xi$. Therefore, our upper bound on $m_{Z'}$ shown in Fig.~\ref{fig:summary} changes only a few percent even in the small coupling region.

\subsection{Supernova Cooling and Neutrino Bursts}
\label{subsec:SNcooling}
The presence of a new MeV force carrier between neutrinos also has interesting implications for the physics of supernova neutrinos. We first briefly summarize the basic picture in the standard case, see Refs.\,\cite{Janka:2006fh, Scholberg:2012id} for review and references therein. The core-collapse supernova forms a proto-neutron star in its core. Its size and temperature are $R \sim 10\,{\rm km}$ and $T \sim 30\,{\rm MeV}$, respectively. In the core, nuclear reactions and electron pair-annihilations produce large numbers of neutrinos. These neutrinos reach thermal equilibrium with nuclear matter and cannot escape from the core due to its high density. As density and temperature decrease with distance from the core, the mean free path of neutrinos becomes longer. Above some radius (called the neutrino sphere), they start streaming freely. The radius and temperature of the $\nu_\mu$ and $\nu_\tau$ sphere are roughly $R \sim 15 \,{\rm km}$ and $T \sim 8 \,{\rm MeV}$, respectively. The $\nu_e$ sphere has a lager size and lower temperature since they can interact with circumstellar media more strongly through the charged current. The diffusion time can be estimated as $\tau_{\rm diff} = \lambda / c (R/\lambda)^{2}$, where $\lambda$ is the neutrino mean free path, and $c$ is the speed of light. If neutrinos have only the SM weak interaction, we can estimate $\tau_{\rm diff}\sim 10\,{\rm s}$, which is consistent with observed duration of the neutrino burst from SN1987A~\cite{Hirata:1987hu, Bionta:1987qt}.

\begin{figure}[t]
\begin{center}
\includegraphics[width=0.9\linewidth]{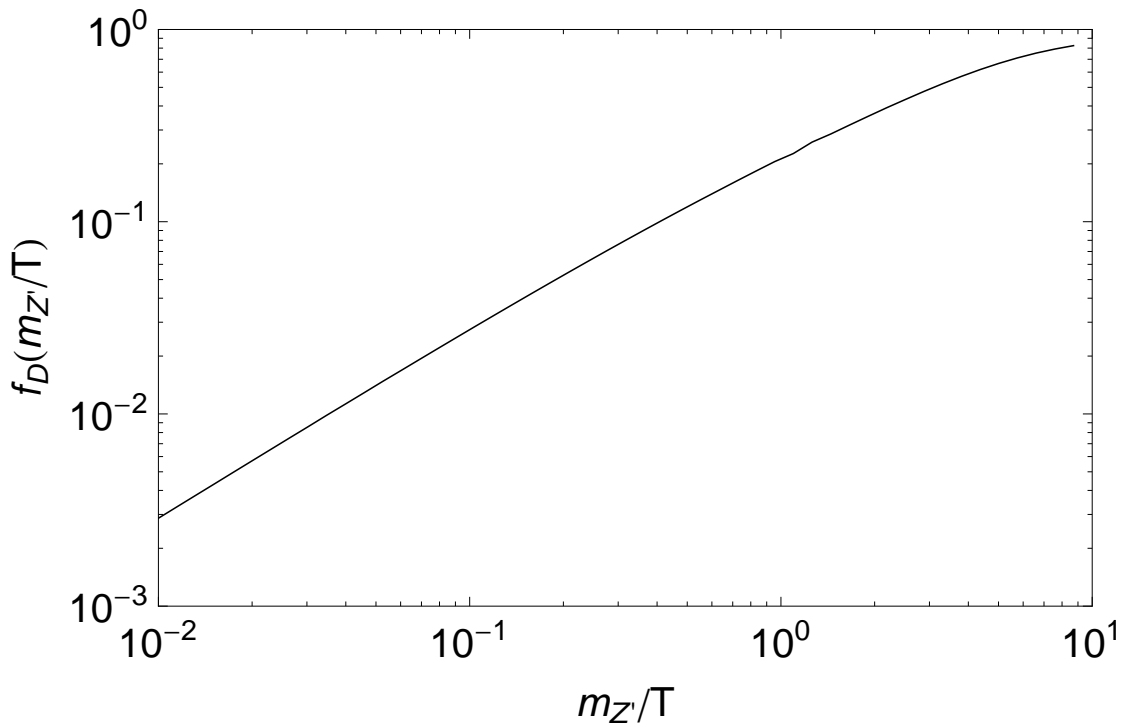}
\caption{\sl \small
Prefactor $f_D$ of the thermally-averaged decay width, $\langle \Gamma_{Z^{\prime}} \rangle = g^{\prime 2} / (12 \pi) m_{Z^{\prime}} f_{D}(m_{Z^{\prime}}/T)$, as a function of the ratio of $Z'$ mass to temperature. By using this result, we evaluate the mean free path of $\nu_\mu$ and $\nu_\tau$ in the supernova core (see Eqs.~(\ref{eq:meanfreepath}) and~(\ref{eq:fd})) .}
\label{fig:fD}
\end{center}
\end{figure}

If neutrinos have new interactions, the standard picture of supernova neutrinos changes. For the $L_{\mu} - L_{\tau}$ model we consider, the $Z'$ mediator can be produced inside the core if its mass is comparable or less than the core temperature. The the travel distance before it decays is only $c \tau_{Z^{\prime}} \sim 10^{-9} \,{\rm km} (g^{\prime} / 10^{-4})^{-2}(T / 10\,{\rm MeV})(10\,{\rm MeV} / m_{Z^{\prime}})^{2}$, which is much smaller than the core radius. Therefore, the produced $Z'$ boson will reach thermal equilibrium with neutrinos and other particles in the core. These reactions may prevent $\nu_\nu$ and $\nu_\tau$ from free-streaming, and change their diffusion time. We evaluate the diffusion time in the following way. The number density of $\nu_\mu$ and $\nu_\tau$ increases as $Z^{\prime}$ decays, ${\dot n}_{\nu} \sim1/2 \langle \Gamma_{Z^{\prime}} \rangle n_{Z^{\prime}}$, where $\langle \Gamma_{Z^{\prime}} \rangle$ is the total decay width of $Z'$ averaged with phase space distributions. Meanwhile, $n_\nu$ decreases through the inverse decay. Therefore, we have ${\dot n}_{\nu} =1/2 \langle \Gamma_{Z^{\prime}} \rangle n_{Z^{\prime}}-\left<\sigma v\right>_{\rm inv}n_\nu n_{\bar{\nu}}$. Since ${\dot n}_\nu=0$ in equilibrium, the detailed balance tells us $\left<\sigma v\right>=n^{\rm eq}_{Z'}/(2\left<\Gamma_{Z'}\right>n^{\rm eq}_\nu n^{\rm eq}_{\bar{\nu}})$, where $n^{\rm eq}$ are the number densities when the particles are in thermal equilibrium as given in Eq.~(\ref{eq:nfornu}). Therefore, we can evaluate the mean-free path for $\nu_\mu$ and $\nu_\tau$ as
\begin{eqnarray}
\label{eq:meanfreepath}
\lambda = \frac{c}{(1/2 \langle \Gamma_{Z^{\prime}} \rangle n^{\rm eq}_{Z^{\prime}} / n^{\rm eq}_{\nu})}.
\end{eqnarray}
To estimate the mean free path in Eq.~(\ref{eq:meanfreepath}), we factorize the thermally-averaged total width as $\langle \Gamma_{Z^{\prime}} \rangle = g^{\prime 2} / (12 \pi) m_{Z^{\prime}} f_{D}(m_{Z^{\prime}}/T)$, where
\begin{eqnarray}
\label{eq:fd}
f_{D} = 16\pi \int \Pi({\vec k}) f_{Z'}(k) \int \Pi({\vec p}) \left(1-f_{\nu}(p)\right) \int \Pi({\vec p'}) \left(1-f_{{\bar \nu}}(p')\right) 
(2 \pi)^{4} \delta^{4} (k - p - p') \,
\end{eqnarray}
with $\Pi({\vec k})=d^{3}{\vec k}/2\omega(2\pi)^{3}$ as the phase space measure. We evaluate $f_{D}$ numerically as shown in Fig.\,\ref{fig:fD}. Roughly speaking, $f_D$ can be regarded as the time dilation factor $f_D\sim m_{Z'}/T$.

\begin{figure}[t]
\begin{center}
\includegraphics[width=0.9\linewidth]{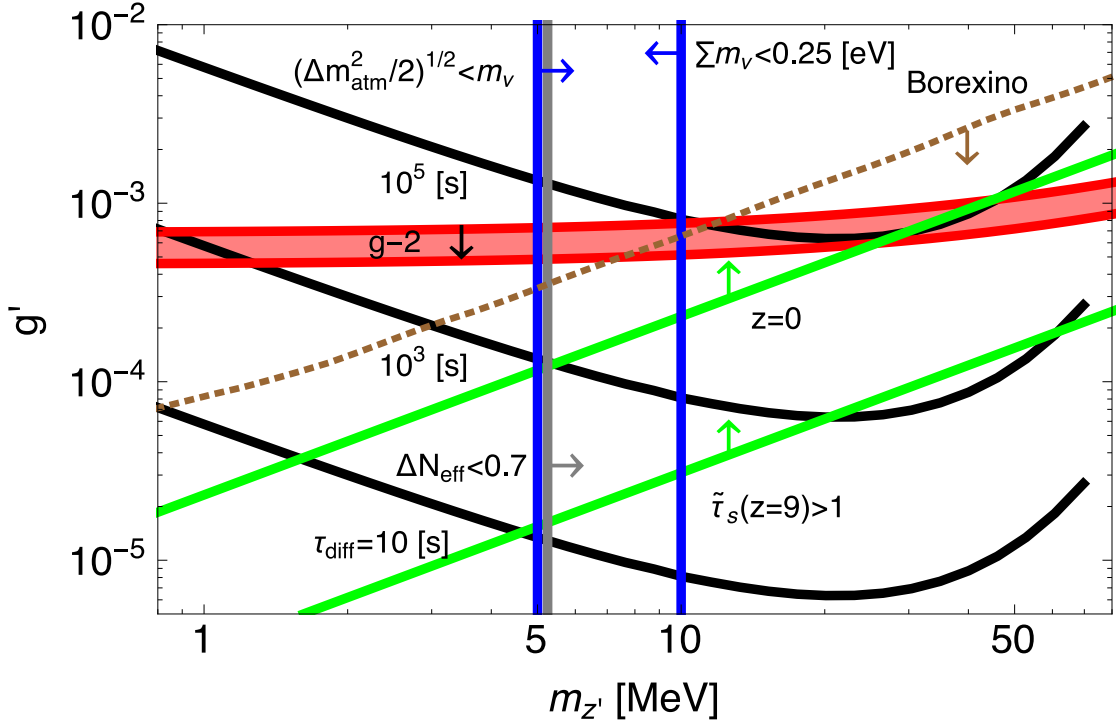}
\caption{\sl \small
Summary of the parameter space of the gauged $L_{\mu}-L_{\tau}$ model. Experimental bounds on the model are from the muon $g-2$ measurement (red band) and the electron-neutrino interaction measurement (brown dotted) at Borexino experiment (Sec.\,\ref{sec:lmu-ltaumodel}). In the case of the quasi-degenerate neutrino mass spectrum, lower and upper bounds (blue vertical) on neutrino masses indicate the range of gauge boson mass $m_{Z'}=5$-$10$\,{\rm MeV} required to produce the IceCube dip via resonant interaction with $C\nu B$ neutrinos (Sec.\,\ref{sec:lmu-ltaumodel}). The optical depth (green) of the IceCube neutrinos ${\tilde \tau}_{s}$ should be larger than unity to reproduce the observed PeV neutrino spectrum (Sec\,\ref{sec:neuprop}). Depending on parameters, resonant scattering can occur at different redshifts. Cosmological constraints requires $m_{Z'}\gtrsim5~{\rm MeV}$ such that ${\mit \Delta} N_{\rm eff} <0.7$ (gray vertical). For $m_{Z'}$ in the range of 5-10 MeV, the energy density carried by the $Z'$ boson may still give sizable contributions to ${\mit \Delta} N_{\rm eff}$, {\it i.e.}, ${\mit \Delta} N_{\rm eff} \sim$0.1-0.7 (Sec.~\ref{subsec:deltaNeff}). The resonant interaction may change the diffusion time of $\nu_\mu$ and $\nu_\tau$ (black). If $\tau_{\rm diff}$ is larger than $\sim$ 10 s, it may delay supernova cooling (Sec.~\ref{subsec:SNcooling}).
}
\label{fig:summary}
\end{center}
\end{figure}

Using Eq.~(\ref{eq:meanfreepath}), we estimate the diffusion time for $\nu_\mu$ and $\nu_\tau$ in the presence of $Z'$ as shown in Fig.~\ref{fig:summary} (solid black), where we have taken the core size as 10 km and temperature 8 MeV.  For $m_{Z'}\sim$5-10 MeV and $g'>10^{-5}$, $\nu_\mu$ and $\nu_\tau$ may not contribute to the neutrino cooling of the core, and the neutrino burst would last $3$ times longer than expected in the standard case. This appears to be incompatible with the observed $\sim 10\,{\rm s}$ duration of the neutrino burst of SN1987A, although we can not draw a concrete conclusion because uncertainties in supernova modeling (e.g., nuclear equation of state), limited statistics of observed events, and also uncertainties in deriving our limit. On the other hand, cooling processes through other invisible particles can compensate the suppressed neutrino cooling in this model. For example, QCD axions with mass of $\sim {\rm meV}$~\cite{Turner:1989vc, Raffelt:1990yz, Raffelt:1999tx} and hidden photons with a mixing parameter of $\sim 10^{-10}$~\cite{Dreiner:2013mua, Burrows:1988ah} are well-motivated candidates in charge of invisible cooling. Running simulations with this model are warranted for improving the limit and comparing it with observations in detail.\footnote{The simulations are performed for QCD axions.~\cite{Burrows:1988ah, Burrows:1990pk}.}

\section{Conclusions}
\label{sec:summary}
The IceCube experiment has observed high-energy cosmic neutrinos for the first time. The observed neutrino spectrum exhibits a dip around the sub-PeV energy scale, which may indicate new physics beyond the SM of particle physics. One possible explanation is that PeV neutrinos may scatter with the \cnb through a MeV resonance and lose their energy before reaching the IceCube detector. In this paper, we have developed a formalism to trace the propagation of PeV neutrinos in the presence of the new interaction. For the flavored interaction, we have shown that resonant scattering may not suppress the PeV neutrino flux completely, which could be tested in the near future.

We have also discussed astrophysical and cosmological constraints on this type of models. The MeV mediator could be produced in the core of supernova, and frequent neutrino collisions induced by the mediator in the core could trap neutrinos inside the core. In the early universe, the mediator could thermalize with the SM thermal bath and contribute to the number of effectively massless degrees of freedom. We have shown both the BBN and supernova observations are sensitive to the favored parameter region explaining the dip in the IceCube high-energy neutrino spectrum.

\section*{Acknowledgments}
We would like to thank Shunsaku Horiuchi, Masahiro Ibe, Joe Sato, Philip Tanedo, and Yue Zhang for helpful discussions. We also thank Joachim Kopp for comments and pointing out an error in an earlier version of the draft. HBY is supported in part by the U. S. Department of Energy under Grant No. DE-SC0008541.

\newpage
\section*{Appendix}

The collision term ${\mathscr C}[{\mathscr F}]$ describes scattering between the PeV neutrinos with \cnb neutrinos, $\nu_{\ell} (k) + \nu_{B \ell^{\prime\prime}} (p) \leftrightarrow \nu_{\ell^{\prime}} (k^{\prime}) + \nu_{B^{\prime} \ell^{\prime\prime\prime}} (p^{\prime}) $, which is given by
\begin{eqnarray}
\label{eq:collision}
{\mathscr C}[{\mathscr F}] 
=&&
-\frac{1}{2\omega(k)} \int d\Pi({\vec k}^{\prime}) \int d\Pi({\vec p}) \int d\Pi({\vec p}^{\prime}) 
(2 \pi)^{4} \delta^{4} (k +p - k^{\prime} - p^{\prime}) 
\sum_{\rm spins} \notag \\
&& \times \frac{1}{2} \sum_{b, b^{\prime}} \Big[ \{ {\mathscr F}({\vec k}, t), {\mathscr M}^{b^{\prime} b \dagger}_{s} \big( 1 - {\mathscr F}({\vec k}^{\prime}, t) \big) {\mathscr M}^{b^{\prime} b}_{s} \} f_{b}({\vec p},t) \left( 1 - f_{b^{\prime}}({\vec p}^{\prime},t) \right) \notag \\
&& - \{ 1 - {\mathscr F}({\vec k}, t), {\mathscr M}^{b^{\prime} b \dagger}_{s} {\mathscr F}({\vec k}^{\prime}, t) {\mathscr M}^{b^{\prime} b}_{s} \} \left( 1 - f_{b}({\vec p},t) \right) f_{b^{\prime}}({\vec p}^{\prime},t)  \Big],
\end{eqnarray}
where $\Pi({\vec k})=d^{3}{\vec k}/(2\omega(k)(2\pi)^{3})$ is the phase space measure, $\{ \cdot, \cdot \}$ denotes the anticommutator, ${\mathscr M}^{b^{\prime} b}_{s, \ell^{\prime} \ell}$ is the scattering amplitude. In deriving Eq.~(\ref{eq:collision}), we have taken a basis in which the density matrix of \cnb neutrinos is diagonal, ${\mathscr F}^{b b^{\prime}}_{B}({\vec k}, t)=\delta^{b b^{\prime}} f_{b}({\vec k},t)$.

The collision term ${\mathscr C}[{\mathscr F}]$ can be further simplified. Since the phase density of the PeV neutrinos is much less than the quantum limit, the Pauli blocking effect is negligible, {\it i.e.}, $1 - {\mathscr F}({\vec k}^{\prime}, t)\simeq1$. We also neglect the inverse scattering process. In addition, we assume the distribution of background neutrinos are flavor-blind, {\it i.e.}, $f_{b}({\vec k}, t) = f({\vec k},t)$, where $f({\vec k},t) = 1/[\exp(k/T_{\nu}(t))+1]$ with $T_{\nu}(t)=T_{\nu,0}(1+z) \simeq 1.7 \times 10^{-4}(1+z)\,{\rm eV}$ and $z$ parametrizing the redshift. With these considerations, the collision term can be written as
\begin{eqnarray}
\nonumber{\mathscr C}[{\mathscr F}] 
= 
&&-\frac{1}{2\omega(k)}\int d\Pi({\vec k}^{\prime}) \int d\Pi({\vec p}) \int d\Pi({\vec p}^{\prime}) 
(2 \pi)^{4} \delta^{4} (k +p - k^{\prime} - p^{\prime})\\&&\times 
\sum_{\rm spins} \frac{1}{2} \{ {\mathscr F}({\vec k}, t), \sum_{b, b^{\prime}} {\mathscr M}^{b^{\prime} b \dagger}_{s} {\mathscr M}^{b^{\prime} b}_{s} \} f({\vec p},t).
\end{eqnarray}
In general, all $s$, $t$ and $u$-channel exchanges of $Z^{\prime}$ contribute to neutrino scattering. Here, we focus on the parameter region where the $s$-channel resonance has a dominant contribution. In this case, the invariant amplitude matrix is
\begin{eqnarray}
{\mathscr M}^{b^{\prime} b}_{s} 
=
{\cal M}\left( \nu (k) + {\bar \nu} (p) \rightarrow \nu (k^{\prime}) + {\bar \nu} (p^{\prime}) \right) {\mathscr O}^{b^{\prime} b} \,,
\end{eqnarray}
with $\mathscr O^{b^{\prime} b}_{\ell^{\prime} \ell} = {\mathscr Q}_{b \ell} {\mathscr Q}_{b^{\prime} \ell^{\prime}}$ (see Eq.\,(\ref{eq:lagrangian})). Hence, we have
\begin{eqnarray}
\sum_{b, b^{\prime}} {\mathscr M}^{b^{\prime} b \dagger}_{s} {\mathscr M}^{b^{\prime} b}_{s}
=2|{\cal M}\left( \nu (k) + {\bar \nu} (p) \rightarrow \nu (k^{\prime}) + {\bar \nu} (p^{\prime}) \right)|^{2} {\mathscr R} \,,
\end{eqnarray}
where ${\mathscr R} = {\mathscr Q}^{\prime \dagger} {\mathscr Q}$. Note that ${\mathscr R}$ contains all information about flavor structure of new neutrino interactions. For the $L_\nu-L_\tau$ model we consider, $\mathscr {\mathscr R}=\diag(0,1,1)$ in the interaction basis. 
Noting the definition of scattering cross section
\begin{eqnarray}
\nonumber 2\omega(k) 2\omega(p) \sigma(\nu (k) + {\bar \nu} (p) \rightarrow \nu + {\bar \nu}) v_{\rm rel}=&& \int d\Pi({\vec k}^{\prime}) \int d\Pi({\vec p}^{\prime}) (2 \pi)^{4} \delta^{4} (k +p - k^{\prime} - p^{\prime})\\
&& \times\sum_{\rm spins}|{\cal M}\left( \nu (k) + {\bar \nu} (p) \rightarrow \nu (k^{\prime}) + {\bar \nu} (p^{\prime}) \right)|^{2} \,,
\end{eqnarray}

we write the scattering rate $\Gamma_{s}({\vec k}, t)$ as 
\begin{eqnarray}
\Gamma_{s}({\vec k}, t)=2 \int \frac{d^{3} {\vec p}}{(2\pi)^{3}} f({\vec p},t) 
\sigma(\nu (k) + {\bar \nu} (p) \rightarrow \nu + {\bar \nu}) v_{\rm rel} \,.
\end{eqnarray}

The cross section for resonant scattering is given by the Breit-Wigner formula
\begin{eqnarray}
\label{eq:bw}
\sigma_{R} = 4\pi \frac{2J+1}{(2s_{1}+1)(2s_{2}+1)} {\rm Br_{in}} {\rm Br_{out}} \frac{1}{p_{\rm cm}^{2}} \frac{E_{\rm cm}^{2} \Gamma_{R}^{2}(E_{\rm cm})}{(E_{\rm cm}^{2}-m_{R}^{2})^{2}+E_{\rm cm}^{2} \Gamma_{R}^{2}(E_{\rm cm})},
\end{eqnarray}
where $s_1$ and $s_2$ are the spins of initial particles, $J$ is the spin of the resonance, $m_R$ is its mass, $\Gamma_R$ is its decay width, and $Br_{\rm in}$ and $Br_{\rm out}$ are decay branching ratios to initial and final state particles, respectively. In the limit of $m_R\gg\Gamma_R$, Eq.~(\ref{eq:bw}) can be written as
\begin{eqnarray}
\label{eq:breightwigner}
\sigma_{R} \simeq 16\pi^{2} \frac{2J+1}{(2s_{1}+1)(2s_{2}+1)} {\rm Br_{\rm in}} {\rm Br_{\rm out}} \frac{\Gamma_{R}}{m_{R}} \delta(E_{\rm cm}^{2}-m_{R}^{2}).
\end{eqnarray}
For the model we consider, $s_1=s_2=0$, ${\rm Br_{\rm in}=Br_{\rm out}}=1/2$, and we have
\begin{eqnarray}
\sigma=12\pi^2\frac{\Gamma_{Z'}}{m_{Z'}}\delta(E^2_{cm}-m^2_{Z'}),
\end{eqnarray}
where $\Gamma_{Z'}=g'^2m_{Z'}/(12\pi)$ is the $Z'$ decay width in the rest frame. Therefore, the total scattering rate is
\begin{eqnarray}
\Gamma_{s}({\vec k}, t)
= 9 \zeta(3) T_{\nu}^{3}(t) \frac{1}{m_{\nu}} \frac{\Gamma_{Z^{\prime}}}{m_{Z^{\prime}}} \delta\left[k-m_{Z^{\prime}}^{2}/(2m_{\nu})\right] \,,
\end{eqnarray}
where $\zeta(s)$ is Riemann zeta function.

\bibliography{ref}

\end{document}